\documentclass[preprint]{aastex}

\usepackage{natbib}
\bibpunct{(}{)}{;}{a}{}{,}                        

\shorttitle{DONUTS: A science frame autoguiding algorithm with sub-pixel precision}
\shortauthors{McCormac et al.}

\begin{document}

\title{DONUTS: A science frame autoguiding algorithm with sub-pixel \\ precision, capable of guiding on defocused stars}

\author{J. McCormac\altaffilmark{1,2}}
\affil{Isaac Newton Group of Telescopes, Apartado de Correos 321, E-38700, Santa Cruz de la Palma, Spain}
\email{jmcc@ing.iac.es}
\and
\author{D. Pollacco}
\affil{Department of Physics, University of Warwick, Coventry CV4 7AL, UK}
\and
\author{I. Skillen}
\affil{Isaac Newton Group of Telescopes, Apartado de Correos 321, E-38700, Santa Cruz de la Palma, Spain}
\and 
\author{F. Faedi}
\affil{Department of Physics, University of Warwick, Coventry CV4 7AL, UK}
\and
\author{I. Todd}
\affil{Astrophysics Research Centre, School of Mathematics and  Physics, Queen's University, Belfast, BT7 1NN, UK}
\and 
\author{C. A. Watson}
\affil{Astrophysics Research Centre, School of Mathematics and  Physics, Queen's University, Belfast, BT7 1NN, UK}

\altaffiltext{1}{Department of Physics, University of Warwick, Coventry CV4 7AL, UK}
\altaffiltext{2}{Astrophysics Research Centre, School of Mathematics and  Physics, Queen's University, Belfast, BT7 1NN, UK}

\begin{abstract}
We present the DONUTS autoguiding algorithm, designed to fix stellar positions at the sub-pixel level for high-cadence time-series photometry, which is also capable of autoguiding on defocused stars. DONUTS was designed to calculate guide corrections from a series of science images and re-centre telescope pointing between each exposure. The algorithm has the unique ability of calculating guide corrections from under-sampled to heavily defocused point spread functions. We present the case for why such an algorithm is important for high precision photometry and give our results from off and on-sky testing. We discuss the limitations of DONUTS and the facilities where it soon will be deployed.  
\end{abstract}


\section{Introduction}
\label{sec:Introduction}

Many areas of modern astronomy require evermore precise time-series photometry. Examples of such areas include: determining the rotation rates of stars, measuring irregularities in asteroid light curves, characterising stellar flares, pulsations and variability and the characterisation of transiting exoplanets and eclipsing binaries. In order to achieve high-precision, potentially sub-millimagnitude (mmag), photometry of bright (photon noise limited) stars, systematic noise in the data must be minimised. As tracking errors and mechanical flexure cause stars to drift around on a CCD they sample different pixels over time. Additional systematic noise may be introduced by uncorrected pixel-to-pixel sensitivity variations, errors from flat fielding, vignetting and changes in seeing. This potentially limits the achievable accuracy of the photometry. In un-guided observations such drifts are inevitable. Even in autoguided observations mechanical flexure and differences in atmospheric refraction between the autoguiding and science cameras will cause stellar images to drift on the CCD.

If the positions of stars can be held fixed in place on the detector throughout a time series of observations, the effects of most of the systematic errors previously mentioned may be reduced or eliminated. However, this is not always possible. Theoretically the best point spread function (PSF) for accurate photometry is a flat topped, top hat-like shape \citep{Howell00}. Orthogonal Transfer (OT) CCDs have the ability to shift their collected charge during an integration to create such a PSF. \citet{2003PASP..115.1340H} investigate the potential use of OT CCDs for high precision photometry. A more commonly used technique is to defocus the telescope (e.g. \citealt{Charbonneau:2000p114,2006ApJ...636..445C,2009MNRAS.396.1023S,2009MNRAS.399..287S,2010MNRAS.408.1680S,2011MNRAS.416.2593B}). Defocusing allows the bulk of the pixels within the PSF to remain the same even in the presence of slight drifts, hence diluting any pixel-to-pixel sensitivity variations. It also limits the effects of changes in seeing as the amount of defocus applied typically dominates the size and shape of the PSF. Finally, defocusing also allows for higher signal-to-noise per unit time observations of bright objects by allowing longer integration times without saturating, simultaneously reducing overheads and increasing efficiency of the duty cycle. 

We present a new autoguiding algorithm, DONUTS, which has been designed to fix telescope pointing to \mbox{$\leq0.2$ pixels} for high-cadence time-series photometry. DONUTS calculates guide corrections directly from the series of science images in real time, essentially eliminating the effects of mechanical flexure and differential refraction between traditional off-axis autoguiders and science imagers. The algorithm is also designed to calculate guide corrections for a range of PSFs, from under-sampled (\mbox{FWHM $\leq 2$ pixels}) to heavily defocused (\mbox{FWHM $\approx 40$ pixels}), so as to maintain all the benefits of defocused photometry that were previously described. Such an algorithm is an essential requirement for the new ground-based transiting exoplanet survey Next Generation Transit Survey\footnotemark\footnotetext{www.ngtransits.org}(NGTS) which aims to push ground-based transit detection limits to the mmag regime to routinely discover Neptune and super Earth-sized exoplanets around bright (V$<13$) stars. 

Transit photometry is crucial in determining the physical properties of exoplanets. Observations of multiple transits allow the orbital period of the planet to be tightly constrained, and modelling the transit's shape allows the orbital inclination and the planet's size relative to its host star to be deduced. When combined with radial velocity measurements of the host star's reflex orbital motion the  \mbox{$\sin\left(i\right)$} ambiguity on the planet's mass is removed and so its bulk properties can be calculated and compared with, or used to constrain, planetary formation and evolutionary models.

This paper is organised as follows, in Section \ref{sec:TheAlgorithm} we describe the DONUTS algorithm used to measure translational shifts in a series of science images. In Section \ref{sec:OffSkyTesting} we present the results of our off-sky testing of DONUTS and in Section \ref{sec:OnSkyTesting} we summarise the results from on-sky testing at the Near Infra-red Transiting Exoplanet Survey (NITES) telescope \citep{2012McCormac_InPrep2}. Finally, in Section \ref{sec:Discussion} we discuss the potential use of DONUTS, comment on its limitations and summarise the work presented here. 

\section{The algorithm and simulated data}
\label{sec:TheAlgorithm}

The typical procedure followed by the DONUTS algorithm in order to provide autoguiding during observations is as follows. After acquiring a target a series of science exposures is started. When the first science image has been read out it is analysed using the method outlined below and becomes the reference image to which subsequent images of this field will be compared to determine guide corrections. When the second science image is acquired it is analysed in a similar manner and pixel shifts in the X and Y directions are calculated. These pixel shifts are then converted to telescope coordinates (e.g. right ascension and declination), sent to the mount (possibly through a control loop, e.g. proportional integral derivative) and the process is repeated for each image acquired. 

For simplicity, in the remainder of this section we will assume that a \mbox{$40\times40$ pixel} image subsection is being used for guiding on a star that has been defocused such that the secondary obstruction is visible in the PSF (hence the name DONUTS). As shown later in Section \ref{sec:OffSkyTesting}, DONUTS is capable of calculating guide corrections from a full CCD array in a fraction of a second, therefore using the positional information from all the stars in the field, while maintaining sub-pixel accuracy. Hence, we envisage that the full frame mode of operation will be used in practice. 

After slewing to a target, a reference image of the field is acquired and analysed by DONUTS. The image is collapsed along the X and Y directions to create two 1-D projections:

\begin{eqnarray}
Image = I \left(x,y \right), \label{eq:Image} \\ 
X_{\mathrm{ref}} = \left(x,\sum_{y}I \left(x,y \right) \right), \label{eq:Xref} \\ 
Y_{\mathrm{ref}} = \left(y,\sum_{x}I \left(x,y \right) \right). \label{eq:Xcomp}
\end{eqnarray}

The projections, $X_{\mathrm{ref}}$ and $Y_{\mathrm{ref}}$, are saved in memory and used as references for subsequent images of this field. On acquiring new images this process is repeated and two comparison projections are created, $X_{\mathrm{comp}}$ and $Y_{\mathrm{comp}}$. To determine the linear shifts in $X$ and $Y$ the corresponding reference and comparison projections are cross correlated using Fast Fourier Transforms \mbox{(FFTs)}. This method was adopted here as the cross correlation of two sets of two 1-D projections was found to be faster than cross correlating two 2-D images. One of the key advantages of autoguiding with DONUTS is that we avoid determining the centroid of our individual guide stars which is difficult for both under sampled and heavily defocused PSFs.

In $X$, we take the FFT of both the reference and comparison projections:

\begin{eqnarray}
FFT_{\mathrm{X_{ref}}} &=& FFT \left(X_{\mathrm{ref}} \right), \label{eq:FFT_X_ref} \\
FFT_{\mathrm{X_{comp}}} &=& FFT \left(X_{\mathrm{comp}} \right). \label{eq:FFT_X_comp}
\end{eqnarray}

\noindent The complex conjugate of the FFT of the comparison projection is then multiplied by the FFT of the reference projection:

\begin{equation}
\centering
\Phi_{\mathrm{X}} = FFT^{\mathrm{*}} \left(X_{\mathrm{comp}} \right) \times FFT_{\mathrm{X_{ref}}}. \label{eq:FFT_Convolution_X} 
\end{equation}

\noindent We then take the Inverse Fast Fourier Transform (IFFT) $\phi_{\mathrm{X}}$ of the previous cross correlation $\Phi_{\mathrm{X}}$ and search the array for the location of the maximum $\phi_{\mathrm{X_{max}}}$:

\begin{equation}
\centering
\phi_{\mathrm{X}} = IFFT \left(\Phi_{\mathrm{X}} \right). \label{eq:FFT_Convolution_X2} 
\end{equation}

\noindent The location, $loc_{\mathrm{X}}$, of $\phi_{\mathrm{X_{max}}}$ in the array of $\phi_{\mathrm{X}}$ is directly related to the shift in pixels between the two projections analysed. Figure \ref{figure:Simulations20121118} shows example guide corrections calculated for simulated X projections of a slightly defocused (such that the simulated PSF becomes double peaked) guide star, both in the case where the overall intensity of the comparison star flux has increased uniformly and where the intensity of the PSF is varying slightly from pixel-to-pixel in the projection. Assuming our reference point as the middle of the CCD array (i.e. pixel 20) for the reference star (see Fig \ref{figure:Simulations20121118} solid red line), $loc_{\mathrm{X}}$ is converted to a guide correction (GC$_{\mathrm{X}}$) using:

\begin{equation}
\mathrm{GC_{X}} \equiv \left\{
\begin{array}{c l}   
   \left( loc_{\mathrm{X}}-40 \right) & \mathrm{when\,\,} loc_{\mathrm{X}} > 20 \\
   loc_{\mathrm{X}} & \mathrm{when\,\,} loc_{\mathrm{X}} < 20 \label{eq:GuideCorrections} \\
   0 & \mathrm{when\,\,} loc_{\mathrm{X}} = 20.
\end{array}\right. 
\end{equation}

The simulations presented in Fig \ref{figure:Simulations20121118} were run \mbox{$10\,000$} times, and all apart from $7$ \mbox{($0.07$\%)} predicted the correct guide correction required to align the reference and comparison projection. Each of the $7$ erroneous predictions were incorrect by a single pixel only. To obtain sub-pixel resolution for real CCD data, quadratic interpolation is performed across the three array elements surrounding $\phi_{\mathrm{X_{max}}}$. This process is simultaneously carried out for the $Y$ direction. We note that in the case of an extremely wide or distorted field a subsection of the CCD array centred on the target of interest would be be used to calculate guide corrections. Working in Fourier space allows quick cross correlation of the two projections created for each CCD axis. The calculation of the image shift is typically returned in a fraction of a second, see Tables \ref{table:1} and  \ref{table:2} for a sample of off-sky guide corrections measured by DONUTS and their corresponding computation time.

\section{Off-sky tests}
\label{sec:OffSkyTesting}

To demonstrate the robustness of DONUTS we tested it off-sky using archive data from four instruments, each with a different PSF, ranging from heavily defocused (FWHM $\approx 40$ pixels) to under sampled (FWHM $\leq 2$ pixels). The entire CCD window was used to calculate guide corrections, therefore using the positional information from all stars in the field, rather than from one star only. The remainder of this section describes briefly the instruments, observations and results from our off-sky testing of DONUTS. Since raw images are used for determining guide corrections in on-sky implementations, we have not reduced the archival images used here. To investigate the effects of reducing images before analysis with DONUTS, we bias subtracted, dark current corrected and flat fielded the test data described in Section \ref{subsec:NGTSGeneva}. We find no improvement in the residual errors from DONUTS using reduced data. Assuming we have previously created a master bias, dark and flat field image, the reduction of each science image before analysis with DONUTS increases the calculation times in Tables \ref{table:1} and \ref{table:2} by at least a factor of $100$. Therefore our assumption of analysing raw data only is justified. 

\begin{figure}
\includegraphics[scale=1.0,trim=20mm 120mm 0mm 0mm]{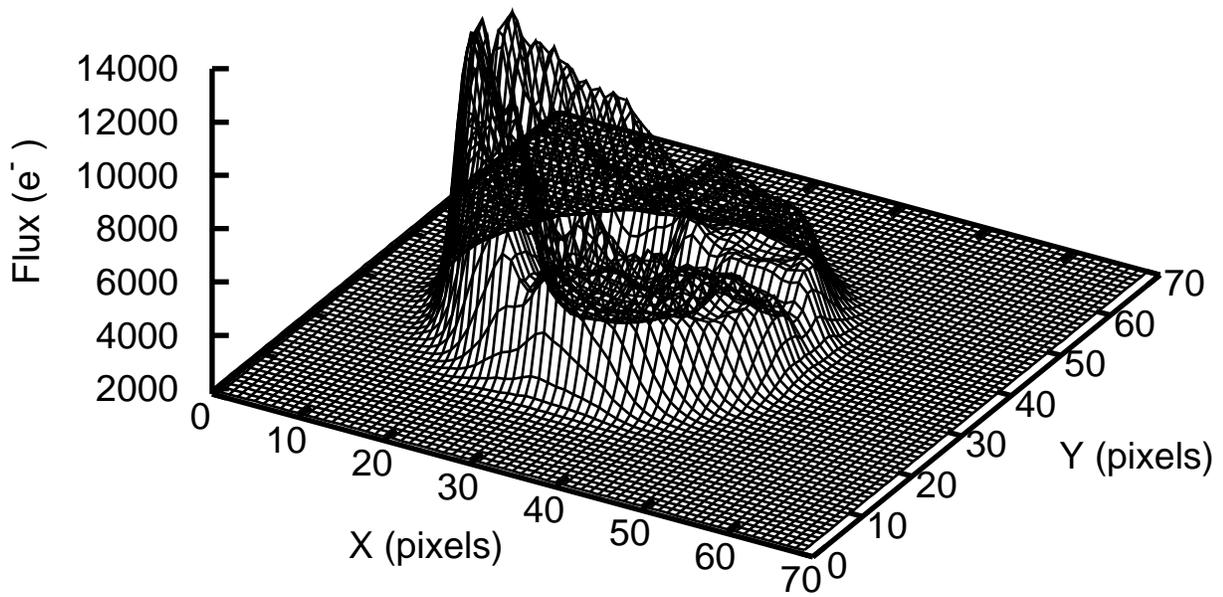}
\caption{Example PSF from the INT/WFC used in off-sky testing.}
\label{figure:int_psf_example}
\end{figure}

\begin{figure}
\includegraphics[scale=1.0,trim=20mm 120mm 0mm 0mm]{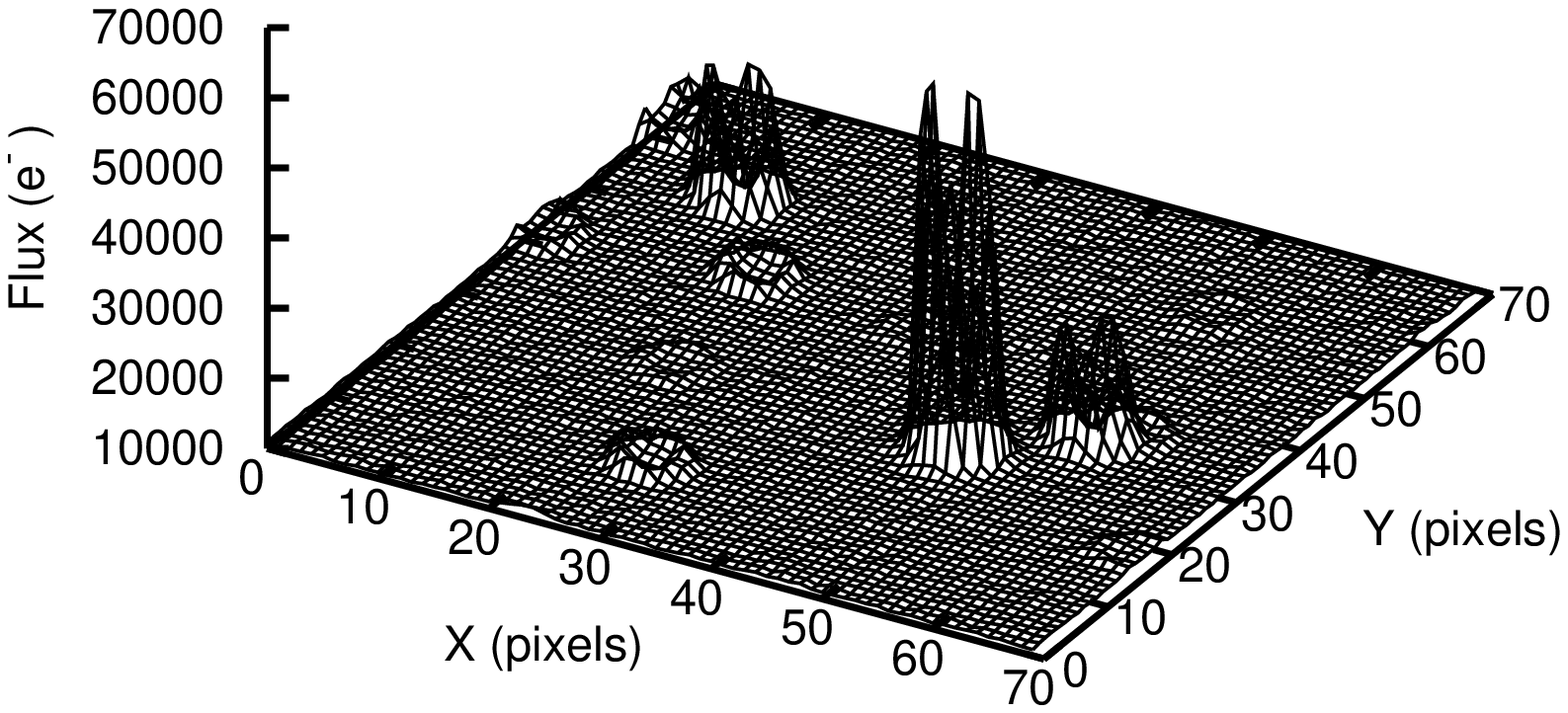}
\caption{Example PSF from the NGTS-P telescope used in off-sky testing.}
\label{figure:ngts_psf_example}
\end{figure}

\begin{figure}
\includegraphics[scale=1.0,trim=20mm 120mm 0mm 0mm]{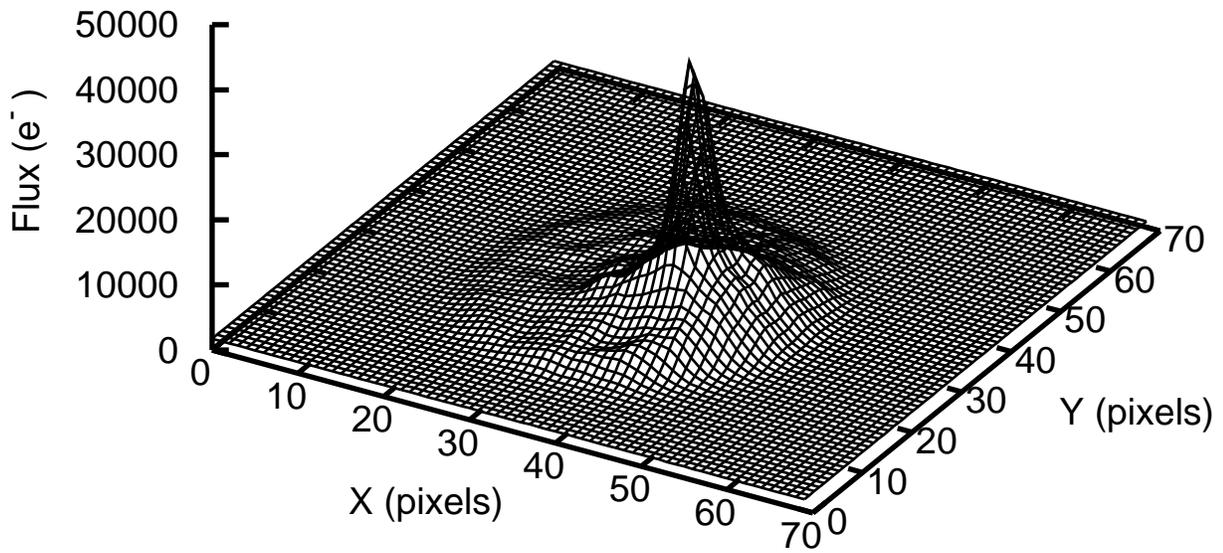}
\caption{Example PSF from the WHT/AG2 used in off-sky testing.}
\label{figure:wht_psf_example}
\end{figure}

\begin{figure}
\includegraphics[scale=1.0,trim=20mm 120mm 0mm 0mm]{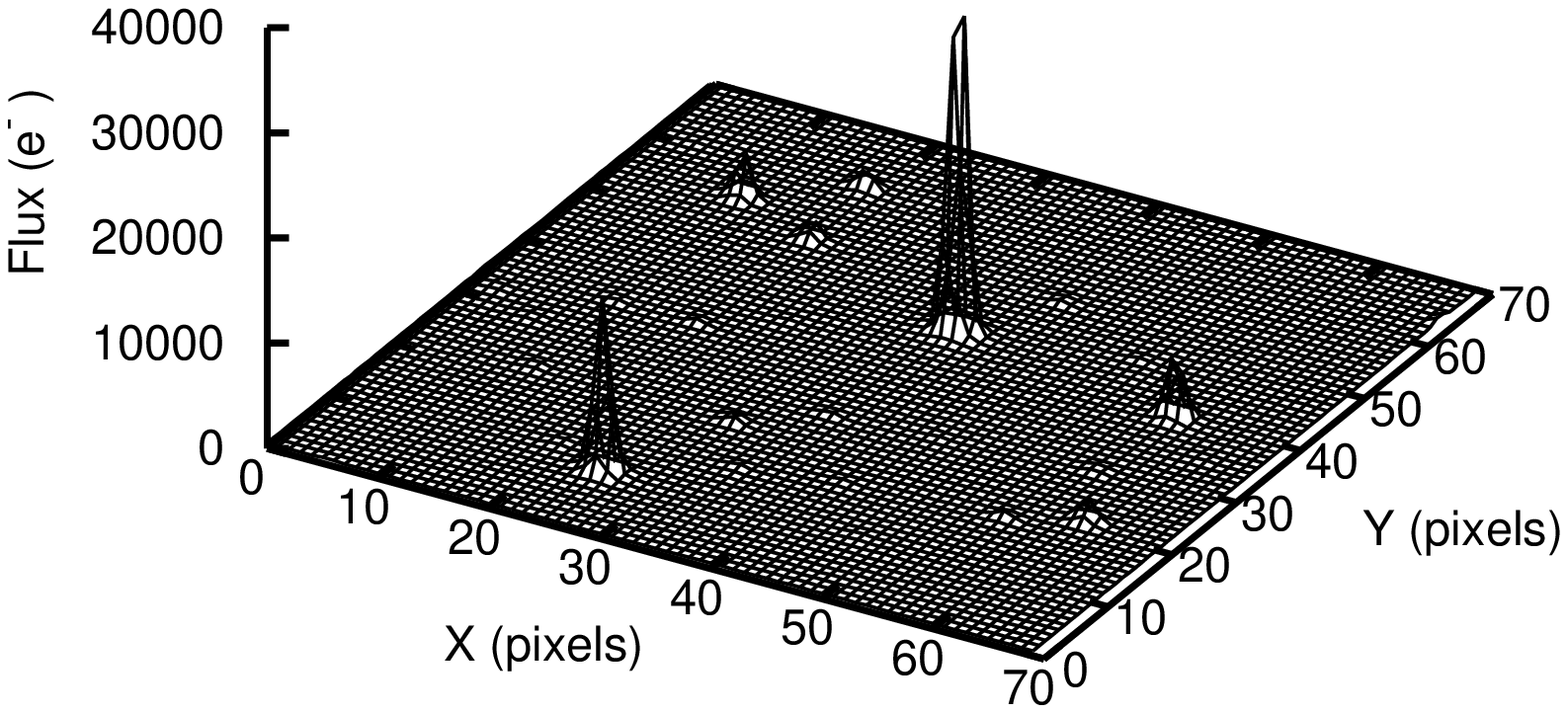}
\caption{Example PSF from the NGTS-1 telescope used in off-sky testing.}
\label{figure:ngts1_psf_example}
\end{figure}

As in initial test, a random image from each instrument described below was chosen as a reference frame and subsequently shifted manually in the X and Y directions $200$ times using IRAF\footnotemark \footnotetext{IRAF is distributed by the National Optical Astronomy Observatories, which are operated by the Association of Universities for Research in Astronomy, Inc., under cooperative agreement with the National Science Foundation.}. The shift magnitudes were chosen randomly between \mbox{$\pm40$ pixels} in each direction. DONUTS was then run on the series of manually shifted images to recover the known shifts applied. A sample of the results from this test for each instrument is given in Table \ref{table:1}. See Figs \ref{figure:int_psf_example}-\ref{figure:ngts1_psf_example} for examples of the PSFs tested off-sky. 

As a second, more realistic, test DONUTS was run once on the complete data series of each instrument below, measuring the required guide correction with respect to the first image in the set. This test is more realistic as the overall intensity of each individual image varies throughout each data set due to changes in airmass and seeing. The images were then shifted by the required correction as determined by DONUTS using IRAF. DONUTS was then run a second time on the corrected data to quantify the residual error. A sample of the guide corrections and their corresponding residuals is given in Table \ref{table:2}. The effects of changes in seeing on DONUTS is investigated in Section \ref{subsec:ChangesInSeeing}.

\subsection{Isaac Newton Telescope - Wide Field Camera}
\label{subsec:INTWFC}

A transit of the SuperWASP \citep{2006PASP..118.1407P} exoplanet candidate \mbox{1SWASP J161732.90+242119.0} was observed on \mbox{2010-05-30} on the Isaac Newton Telescope (INT), with the Wide Field Camera (WFC). To increase the signal-to-noise ratio per unit time and to alleviate saturation the telescope was heavily defocused to \mbox{$\approx40$ pixels} \mbox{($13\farcs2$)} FWHM (see Fig \ref{figure:int_psf_example}). To increase the cadence, the WFC was also windowed from \mbox{$4\mathrm{k}\times2\mathrm{k}$} to \mbox{$2285\times1465$ pixels}. Defocusing the WFC also defocuses the autoguider as the CCDs are in the same focal plane, and so the observations were guided manually, re-centering the telescope pointing every few minutes. A total of $167$ manually guided images were obtained and subjected to the two tests described above. Regions of the CCD containing dead columns were excluded from guiding calculations. Small samples of the off-sky guide measurements and their residuals are given in Tables \ref{table:1} and \ref{table:2}. 

The average residual errors from DONUTS when shifting a random image manually $200$ times were only \mbox{$0.01$ pixel} in each of the X and Y directions. The average residuals measured from the complete data series of $167$ images were $0.02$ and \mbox{$0.01$ pixels} in the X and Y directions, respectively. The pixel size of the WFC is \mbox{$0\farcs33$ pixel$^{-1}$}, therefore the theoretical average error from DONUTS corresponds to $0\farcs007$ and $0\farcs003$ on-sky. This, of course, is much smaller than the finest control of the INT \mbox{($\approx0\farcs1$)} and the real error from DONUTS will most likely be dominated by the telescope's limitations to make small guide corrections.

\subsection{Next Generation Transit Survey - Prototype Telescope}
\label{subsec:NGTS}

On 2010-06-29 a total of $150$ images with exposure times of \mbox{$60$ s} were obtained of the transiting super-Earth host GJ 1214 \citep{2009Natur.462..891C} during photometric testing of the NGTS prototype system (hereafter \mbox{NGTS-P}) in 2009/10. NGTS-P suffered from mechanical flexure between the separate science and autoguiding telescopes as well as imperfect polar alignment. This was seen as a slow drift in stellar positions across the CCD in the Y direction with a slight drift in X \mbox{($3.79$ pixel h$^{\mathrm{-1}}$ total)}. The telescope was defocused to \mbox{$4.8$ pixels} \mbox{($25\farcs7$)} FWHM (see Fig \ref{figure:ngts_psf_example}) to minimise the effects of the drift. The observations of GJ 1214 were analysed with DONUTS using the two methods previously described; small subsets of the off-sky guide corrections and their residuals are given in Tables \ref{table:1} and  \ref{table:2}. 

The average residual errors from DONUTS when shifting a random image manually $200$ times were $0.16$ and \mbox{$0.11$ pixels} in the X and Y directions, respectively. The average residuals over the complete data set of $150$ images were $0.12$ and \mbox{$0.08$ pixels}, also in the X and Y directions, respectively. The plate scale of NGTS-P is \mbox{$5\farcs36$ pixel$^{\mathrm{-1}}$} resulting in on-sky residuals of $0\farcs643$ and $0\farcs429$ in X and Y, respectively. As in Section \ref{subsec:INTWFC}, the average residuals from DONUTS off-sky testing at NGTS-P are also close to the limit of the telescope's repointing capability.  Therefore, as before we suspect that the real error in autoguiding with DONUTS will be dominated by the response of the telescope.

\subsection{William Herschel Telescope - AG2}
\label{subsec:WHTAG2}

AG2 was a temporary instrument with a frame-transfer CCD mounted at the folded Cassegrain focus of the William Herschel Telescope (WHT). It had a Field of View (FOV) of \mbox{$3\farcm3\times3\farcm3$} and a plate scale of \mbox{$0\farcs4$ pixel$^{\mathrm{-1}}$} and was used for transit timing variation surveys of known exoplanets (e.g. \citealt{2010MNRAS.403.2111H}). On the night of 2007-09-19 a transit of WASP 1b \citep{2007MNRAS.375..951C} was observed with \mbox{$10$ s} exposures. The field of AG2 was significantly distorted resulting in an asymmetric PSF with a mean FWHM of \mbox{$5.71$ pixels} (see Fig \ref{figure:wht_psf_example}). AG2 images were cropped before off-sky testing as they contained many pixels that were located outside the beam of the telescope and hence not illuminated. Observations of WASP 1b were classically autoguided with the WHT's Cassegrain autoguider and hence showed very little movement in the stellar positions. Also, due to the small FOV of AG2, only $3$ objects were visible on the CCD. The data were analysed using the methods previously described and samples of the results are given in Tables \ref{table:1} and \ref{table:2}. 

The average residual errors from DONUTS when shifting a random image manually $200$ times were $0.006$ and \mbox{$0.005$ pixels} in the X and Y directions, respectively. The average residuals from the raw data set were $0.002$ and \mbox{$0.006$ pixels} in the X and Y directions, respectively.  This was expected as the stellar positions on the CCD were quite stable over time. As before, DONUTS appears to perform remarkably well regardless of the instrumental PSF, predicting guide corrections accurately at the sub-pixel level. 

\subsection{Next Generation Transit Survey}
\label{subsec:NGTSGeneva}

In August 2012 the first telescope (hereafter NGTS-1) for the full NGTS facility was delivered to the Geneva Observatory, Switzerland for characterisation and testing before being shipped to Paranal, Chile. The optical system of NGTS employs \mbox{$20$ cm} aperture telescopes with fast focal ratios of $f/2.8$ made by Astro Syteme Austria (ASA) and Andor Technologies \mbox{iKon-L}, \mbox{$2\mathrm{k}\times2\mathrm{k}$} CCD cameras with deep-depletion. Each telescope has a FOV and plate scale of \mbox{$2\fdg83\times2\fdg83$} and \mbox{$4\farcs96$ pixel$^{\mathrm{-1}}$}, respectively. On 2012-08-17 a series of $600$ images with \mbox{$4$ s} exposure times was taken during tests of the telescope's tracking performance. The images were focused such that the stars became under sampled with \mbox{$\mathrm{FWHM}<2$ pixels} (see Fig \ref{figure:ngts1_psf_example}). It was found that after precise polar alignment the system was remarkably stable and the stars showed very little long term drifting \mbox{($<1$ pixel h$^{-1}$)} when tracking only. Analysis of NGTS-1 data with DONUTS off-sky using the methods described above also showed that the system required little correction, see Table \ref{table:2} for a sample of the results. 

The average residual errors from DONUTS when shifting a random image manually $200$ times were $0.18$ and \mbox{$0.11$ pixels} in the X and Y directions, respectively. The average guide corrections required to recenter the raw data set were $0.29$ and \mbox{$0.36$ pixels} in the X and Y directions, respectively with average residuals of $0.09$ and \mbox{$0.08$ pixels} in X and Y over the $600$ corrected images analysed. The average residuals correspond to on-sky movements of $\approx0\farcs40$ which is slightly larger than for the instruments previously described due to the larger pixel size, but still close to the limit of the mount's pointing ability. Several individual residual measurements (X$_{\mathrm{r}}$) on the x-axis of NGTS-1 can be seen in Table \ref{table:1} to be slightly elevated. This was caused by an incorrect setup of the CCD clocking voltages which induced streaking of saturated stars in many of the images analysed. This issue has since been resolved and optimisation of DONUTS for the full NGTS facility is ongoing.

\subsection{Limiting signal-to-noise ratio and V magnitude of a single guide star}
\label{subsec:LimSNR}

When autoguiding with DONUTS it is important to quantify the limiting signal-to-noise ratio of the guide stars (including CCD noise) below which guide corrections become unreliable and therefore the algorithm breaks down. To ascertain this worst case scenario we created a series of \mbox{$50\times50$ pixels} image stamps, each containing a single star in focus, with a Gaussian profile \mbox{($\mathrm{FWHM}=2.5$ pixels)}. The signal-to-noise ratio (SNR) of the star was varied between $0.5\leq\mathrm{SNR}\leq20.0$ in steps of $0.5$. At each SNR the image of the artificial star was shifted manually $200$ times in the X and Y directions. The shifts in each direction were randomly selected from values ranging between $-10$ and $10$ pixels, with $0.01$ pixel resolution. The same set of $200$ shifts were used for all SNRs tested. 

We define a stringent, maximum residual error and standard deviation in the residual shifts from DONUTS of $0.3$ pixel. This level of precision was surpassed when the SNR ratio of the manually shifted artificial star reached $\mathrm{SNR}\approx9$. Therefore, we place a conservative limiting $\mathrm{SNR}=15$ for the operation DONUTS. 

\begin{figure}
\includegraphics[scale=0.65, angle=270]{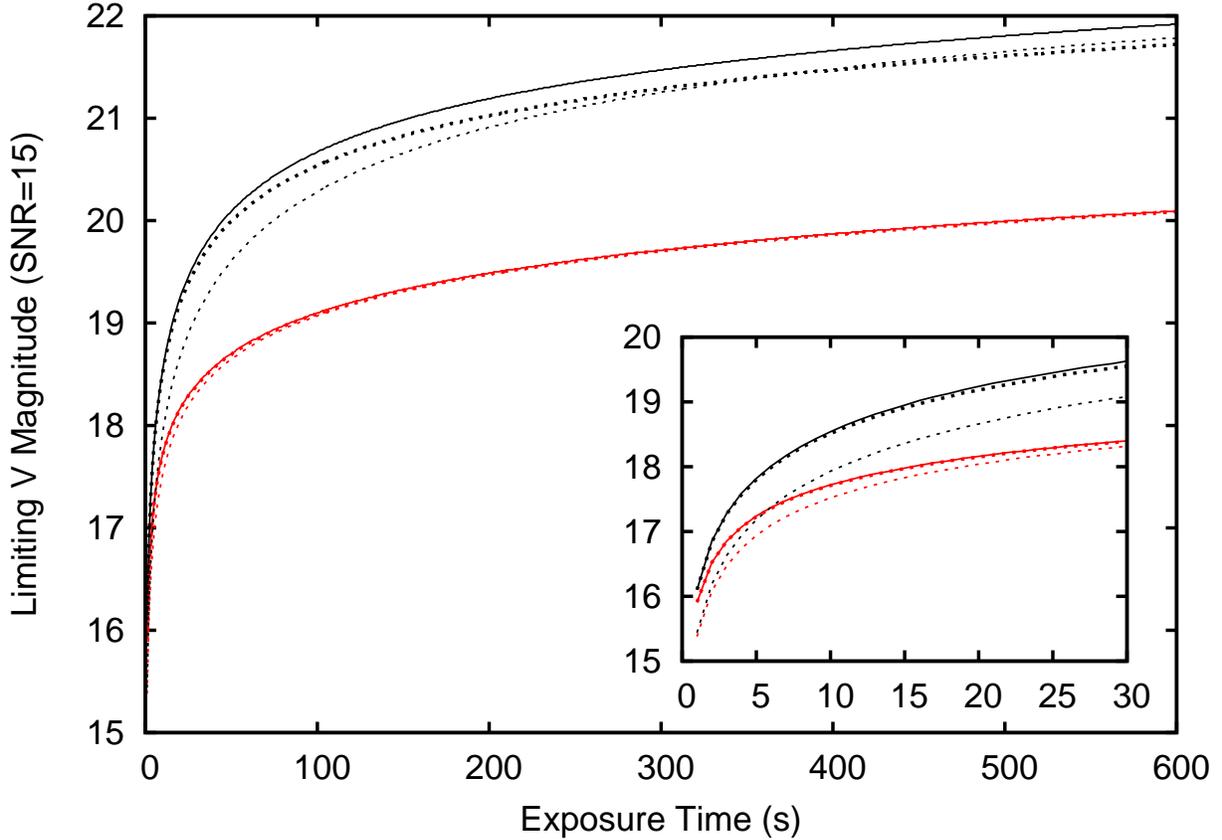}
\caption{Limiting V magnitude of the DONUTS autoguiding algorithm as a function of exposure time for the worst case scenario where only one guide star is available. The solid lines show our baseline model of a $1$ m telescope and assuming the parameters given in Table \ref{table:3}. The dashed lines represent an increase in read noise of $10$ e$^{-}$ pixel$^{-1}$ on top of the baseline model ($20$ e$^{-}$ pixel$^{-1}$ total) and the dotted lines simulate a considerable dark current increase from $0.2$ to $1$ e$^{-}$ s$^{-1}$ pixel$^{-1}$. The red and black contours represent full and new moon observations, respectively and the insert shows a zoom around short exposure times. As expected, increasing the read noise has a strong effect on the limiting guide star brightness at short exposure times. Increasing the level of dark current has a noticeable effect on the limiting guide star brightness at long exposure times, especially during new moon conditions. Increasing the CCD noise during full moon conditions has little effect when $T_{\mathrm{exp}}>30$ s as the noise from the sky begins to dominate.}
\label{figure:LimSNR15}
\end{figure}

In Fig \ref{figure:LimSNR15} we investigate as a function of exposure time the limiting V magnitude of a single guide star needed to give \mbox{$\mathrm{SNR}=15$}. We calculate contours for a $1$ m telescope aperture (these can be scaled to different apertures) for bright and dark sky conditions, and for a range of CCD parameters (see Table \ref{table:3}). We assume a baseline model CCD with a read noise of $10$ e$^{-}$ and dark current of $0.2$ e$^{-}$ s$^{-1}$ pixel$^{-1}$ and investigate the effects of increasing these noise parameters to simulate lower quality CCD cameras. Using Fig \ref{figure:LimSNR15} and,

\begin{equation}
\centering
T_{\mathrm{exp}} \leq T_{\mathrm{c}} - T_{\mathrm{r}}\label{eq:Tc} 
\end{equation}

\noindent we can calculate a range of allowable exposure times ($T_{\mathrm{exp}}$) and guide star brightnesses, where $T_{\mathrm{c}}$ is the maximum allowable dwell time before the pointing of the telescope needs corrected and $T_{\mathrm{r}}$ is the read out time of the CCD. Assuming $T_{\mathrm{c}}=120$ s and $T_{\mathrm{r}}=20$ s the exposure time must satisfy $T_{\mathrm{exp}}\leq 100$ s, resulting in limiting guide star brightnesses of $V\approx19$ and $20.5$ for our baseline model during bright and dark sky conditions, respectively.

CCDs with higher read noise require brighter guide stars to reach the required SNR=15 when observing at high cadence, this is because the read noise from the CCD is significant in comparison to the noise from the guide star and sky background. As the exposure time is increased, the dashed lines of increased read noise in Fig \ref{figure:LimSNR15} tend to converge with the baseline model, this convergence is most pronounced during full moon as the noise from the sky quickly begins to dominate. CCDs with increased dark current require slightly brighter guide stars when observing with long exposures as the flux from dark current builds up quickly. We can see this as a very slight divergence of the dotted lines in Fig \ref{figure:LimSNR15} from the baseline model. However, by far the most dominant factor in determining the limiting magnitude of a guide star, for a given exposure time, is the sky brightness. Figure \ref{figure:LimSNR15} shows a change in limiting V magnitude of approximately $2$ mag between full and new moon. Defocused stars were not analysed in these simulations as by definition defocusing is only applied when the target is bright and therefore does not represent the limiting case of autoguiding low SNR observations with DONUTS. 


\subsection{Changes in seeing}
\label{subsec:ChangesInSeeing}

When observing a field of heavily defocused stars, changes in seeing have very little effect on the effective PSF $\sigma_{\mathrm{PSF}}$ as the focus term in 

\begin{equation}
\centering
\sigma_{\mathrm{PSF}} = \sqrt{\sigma^{2}_{\mathrm{seeing}}+\sigma^{2}_{\mathrm{focus}}}, \label{eq:effectivePSF} 
\end{equation}

\noindent tends to dominate, where $\sigma_{\mathrm{seeing}}$ and $\sigma_{\mathrm{focus}}$ are the uncorrelated FWHMs of the seeing disk and telescope focus in arcsec, respectively. However, when observing close to focus, changes in seeing may dramatically alter the PSF. This is most apparent for instruments whose pixel scale is less than the typical site seeing for the observatory where it is located. 

To test the effects of changes in seeing on DONUTS we conducted the following simulation. We assume that a reference image for a given field has been observed in focus and under good seeing conditions, however on a subsequent observation of the same field the seeing has deteriorated. To simulate the worst case scenario our reference image was blurred with a Gaussian blurring filter and shifted manually $200$ times. As before, the shifts were chosen from a range of \mbox{$\pm40$ pixels} in both the X and Y directions. This process was then repeated for $8$ levels of seeing deterioration, ranging from a $0$\% to $125$\% increase in the average stellar FWHM. 

We applied this method to our under sampled NGTS-1 data, using the same reference image as used in Section \ref{subsec:NGTSGeneva}. The same $200$ shifts applied during our off-sky testing were applied to the reference frame and all but the reference image were blurred using a Gaussian blurring filter to the FWHMs given in Table \ref{table:4}. In a similar manner to the first off-sky test in Section \ref{sec:OffSkyTesting}, DONUTS was then run on the shifted and blurred data and the measured shifts were compared to the known offsets applied. The average residuals for each level of simulated seeing change are compared to the normal data in Table \ref{table:4}. As can be seen in Table \ref{table:4} changes in seeing have only a marginal effect on the residual errors of DONUTS. 

\section{On-sky test}
\label{sec:OnSkyTesting}

\begin{figure}
\includegraphics[scale=0.65,angle=270]{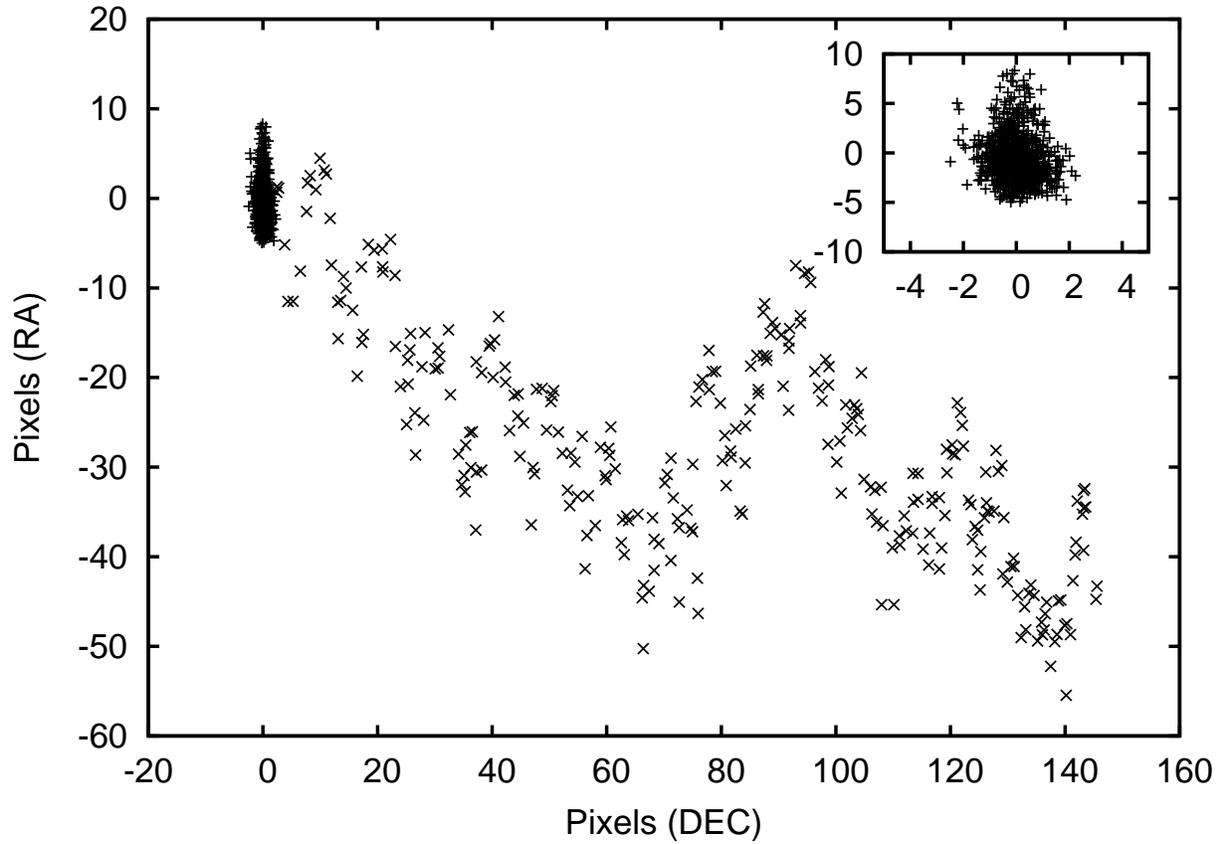}
\caption{Example plots of stellar positions across the NITES CCD with DONUTS disabled (X symbols) and enabled (cross symbols, see subplot insert). The position of each star has been normalised to (0,0) using its position in the first image. Although the NITES telescope suffers badly from periodic error in the RA axis there is a clear improvement in using DONUTS over tracking only.}
\label{figure:donuts_onoff}
\end{figure}

DONUTS was first tested on-sky at the NITES telescope \citep{2012McCormac_InPrep2} on La Palma. The NITES telescope is a remotely operated \mbox{$0.4$ m} Schmidt-Cassegrain telescope with a $1024\times1024$ pixel CCD camera with deep depletion technology. The FOV and plate scale of the telescope are  $11\farcm26\times11\farcm26$ and $0\farcs66$ pixel$^{-1}$. The NITES telescope suffers from periodic error \mbox{($\mathrm{P}\approx4$ min}) in right ascension caused by mechanical inaccuracies in the drive system. During commissioning the periodic error was modelled and corrected, reducing it from approximately $\pm 30$ to \mbox{$\pm 6$ pixels}. As DONUTS applies guide corrections between science images the periodic error essentially limits the maximum exposure time of NITES to \mbox{$t_{\mathrm{max}}=30$ s} without the use of conventional autoguiding. Figure \ref{figure:donuts_onoff} shows the effects of enabling DONUTS at the NITES telescope. Figure \ref{figure:donuts_onoff} (X symbols) shows the drift in position of a star observed on \mbox{2011-02-18} with DONUTS disabled. The observations lasted $3.5$ h and a drift can clearly be seen in declination. This is caused by a slight polar misalignment of the telescope. The vertical spread in the drift profile is caused by the periodic error of the mount. Figure \ref{figure:donuts_onoff} (cross symbols, see subplot insert) shows data taken over $4.3$ h on \mbox{2012-11-19} with DONUTS enabled. A clear improvement over tracking only can be seen. The higher the cadence of observations made with the NITES telescope the more times the periodic error is sampled and corrected by DONUTS. If deployed on a mount with little or no periodic error (e.g. NGTS-1) we expect extremely stable tracking performance over long periods of time.

\subsection{Discussion and summary} 
\label{sec:Discussion}

\begin{figure}
\includegraphics[scale=0.65,angle=270]{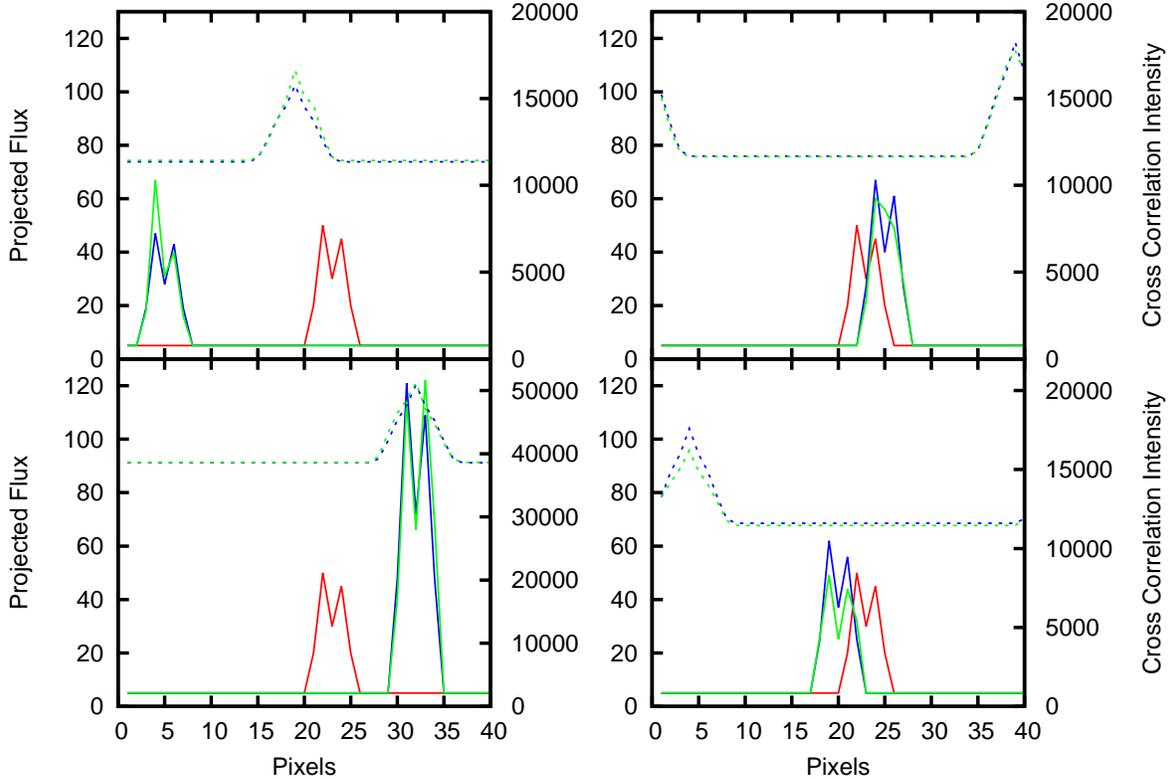}
\caption{Simulated guide star testing with a slight defocus, such that the simulated PSF becomes double peaked. The solid red line represents a \mbox{1-D} projection of a typical guide star in the X direction, where a slight telescope defocus has been applied. The solid blue line represents the same guide star with a uniform change in intensity applied to all pixel elements. The solid green line represents the same as the solid blue line plus a random pixel-by-pixel change in intensity where the random values are selected using a Gaussian distribution with a mean of $0$ and a standard deviation of $1$ from \mbox{$\pm20$\%} of the intensity modified (blue) pixel value. The dashed blue and green lines represent the IFFT, $\phi_{\mathrm{X}}$, from \mbox{Eq \ref{eq:FFT_Convolution_X2}} of the solid blue and green comparison projections above. The location, $loc_{\mathrm{X}}$, of the peak  $\phi_{\mathrm{X_{max}}}$, in $\phi_{\mathrm{X}}$, with respect to the x-axis origin is converted to a guide correction using Eq \ref{eq:GuideCorrections}. The same treatment is also applied to the Y axis projections to calculate the guide correction in the Y direction.}
\label{figure:Simulations20121118}
\end{figure}

Our initial simulations of DONUTS in Section \ref{sec:TheAlgorithm}, using a simulated star which had been slightly defocused returned encouraging results. This test showed that DONUTS is capable of measuring translational shifts between two 1-D projections accurately. Providing the limiting \mbox{$\mathrm{SNR}=15$} is achieved, changing the overall intensity of the comparison projection in Fig \ref{figure:Simulations20121118} has no effect on determining the shift correctly. Also, subsequently tweaking the individual pixels of the projection randomly by a moderate amount, and varying the overall intensity of the comparison frame has no effect on the shift measurement. The simulations in Section \ref{sec:TheAlgorithm} were run $10\,000$ times, with $0.07$\% of the measurements being incorrect by a single pixel only. 

In Section \ref{sec:OffSkyTesting} we demonstrated DONUTS capability of rapidly calculating guide corrections using a full CCD array and a variety of PSFs, while maintaining an average residual error well within the sub-pixel (\mbox{$<0.2$ pixel}) regime. We define a conservative limiting SNR in the worst case scenario of $\mathrm{SNR}=15$ and investigate the effects of changing sky brightness, dark current and read noise on the limiting V magnitude of a single guide star in the general case of a $1$ m telescope. The benefits of spatial stability combined with the ability to eliminate mechanical flexure, while simultaneously defocusing (if desired) will aid astronomers in pushing the limits of photometric accuracy to the sub-millimag level in many areas of astronomy. For example, near infrared observations of faint brown dwarfs or secondary eclipses of transiting exoplanets, which are typically carried out in focus to minimise the effects of the infrared background, could benefit as much from DONUTS as heavily defocused optical observations of bright transiting exoplanet host stars. 

As seen in Table \ref{table:2} several guide corrections with magnitudes \mbox{$\leq0.1$ pixels} become difficult to correct, however at this level of spatial stability we are theoretically already able to meet our goals for DONUTS in providing sub-pixel stability (\mbox{$\leq0.2$ pixels}) for high-cadence time-series photometry. 

Section \ref{sec:OnSkyTesting} clearly highlights the main limitation of the DONUTS autoguiding algorithm. Deployment of DONUTS on a poorer quality mount results in a general localisation of stellar positions rather than the absolute sub-pixel fixing we desire. However, as shown in Section \ref{subsec:NGTSGeneva} the NGTS-1 telescope is extremely stable and we therefore expect unprecedented tracking performance for NGTS when DONUTS is enabled. 

We note that, over wide FOVs, spatial differential refraction caused by the atmosphere may introduce a slight error if the entire FOV is used to calculate guide corrections. Under such conditions the guide corrections may be calculated using a small subsection of the CCD surrounding the target of interest. Our investigation into the effects of changes in seeing on DONUTS also returned promising results. The average PSF in the comparison frames tested was increased by up to \mbox{$125$\%} with almost no effect on the level of residual error from DONUTS. This was expected as the maximum correlation between two Gaussian profiles occurs when their peaks overlap, regardless of the increase in FWHM. As so long as the telescope is not physically defocused by a large amount (such that the secondary obstruction is visible) between acquiring the reference and comparison images we expect DONUTS to handle slight variations in PSF and overall image intensity in a robust manner. 

We have presented a new autoguiding algorithm designed to correct a telescope's pointing between the science images of high-cadence time-series photometry, with the goal of producing stable telescope tracking at the sub-pixel (\mbox{$\leq0.2$ pixel}) level. Our tests have shown that autoguiding with DONUTS will theoretically be superior than conventional autoguiding up to an exposure time where telescope tracking errors become a significant fraction of the PSF size. DONUTS simultaneously eliminates mechanical flexure issues and atmospheric refraction effects as a target is tracked through a range of airmasses. We conclude that the algorithm is significantly robust at handling different PSF sizes and shapes and is robust against changes in seeing. We find an average residual error \mbox{$<0.2$ pixels} over the range of PSFs tested. The main limitation of DONUTS is expected to be the response of a given telescope to small guiding corrections. One possible way of circumventing this issue is to characterise the telescope response and implement a control loop, e.g. PID, to account for over or under correction made by the telescope. Tailored versions of DONUTS are currently under development for the NGTS facility at Paranal in Chile and for the \mbox{$1$ m} SuperWASP Qatar Telescope (SQT) at the Observatorio del Roque de los Muchachos, on the island of La Palma in the Canary Islands.

{\it Facilities: }\facility{NGTS}, \facility{NGTS-P}, \facility{ING:Newton (WFC)}, \facility{ING:Herschel (AG2)}.

\clearpage



\begin{deluxetable}{rrrrrrrrrrrrrrrrrrrr}
\tabletypesize{\tiny}
\addtolength{\tabcolsep}{-5pt}
\rotate
\tablecaption{A small sample of the shifts applied manually to a reference image from each of the $4$ instruments tested. X$_{\mathrm{m}}$ and Y$_{\mathrm{m}}$ are the manually applied shifts in pixels. X$_{\mathrm{rm}}$ and Y$_{\mathrm{rm}}$ are the differences in pixels between the shift applied and that measured by DONUTS and t$_{\mathrm{calc}}$ is the time in ms of the calculation.}
\tablewidth{0pt}
\tablehead{
\multicolumn{5}{c}{INT - WFC} & \multicolumn{5}{c}{NGTS-P} & \multicolumn{5}{c}{WHT - AG2} & \multicolumn{5}{c}{NGTS-1}\\
\colhead{X$_{\mathrm{m}}$} & \colhead{Y$_{\mathrm{m}}$} & \colhead{t$_{\mathrm{calc}}$} & \colhead{X$_{\mathrm{rm}}$} & \colhead{Y$_{\mathrm{rm}}$} & \colhead{X$_{\mathrm{m}}$} & \colhead{Y$_{\mathrm{m}}$} & \colhead{t$_{\mathrm{calc}}$} & \colhead{X$_{\mathrm{rm}}$} & \colhead{Y$_{\mathrm{rm}}$} & \colhead{X$_{\mathrm{m}}$} & \colhead{Y$_{\mathrm{m}}$} & \colhead{t$_{\mathrm{calc}}$} & \colhead{X$_{\mathrm{rm}}$} & \colhead{Y$_{\mathrm{rm}}$} & \colhead{X$_{\mathrm{m}}$} & \colhead{Y$_{\mathrm{m}}$} & \colhead{t$_{\mathrm{calc}}$} & \colhead{X$_{\mathrm{rm}}$} & \colhead{Y$_{\mathrm{rm}}$} \\
\colhead{(pixels)} & \colhead{(pixels)} & \colhead{(ms)} & \colhead{(pixels)} & \colhead{(pixels)} & \colhead{(pixels)} & \colhead{(pixels)} & \colhead{(ms)} & \colhead{(pixels)} & \colhead{(pixels)} & \colhead{(pixels)} & \colhead{(pixels)} & \colhead{(ms)} & \colhead{(pixels)} & \colhead{(pixels)} & \colhead{(pixels)} & \colhead{(pixels)} & \colhead{(ms)} & \colhead{(pixels)} & \colhead{(pixels)} \\ 
}
\startdata
18.64	& 9.46	& 6.4 & -0.01	& 0.00	& 9.74	& -7.58	& 1.4	 & -0.10	& 0.07	& -6.67	& -6.48	& 1.1 & 0.00	& 0.00	& 7.10	& -17.08	& 2.2 & 0.05	& 0.08	\\
-9.11		& -12.87	& 6.2 & 0.01	& 0.00	& -3.66	& -8.07	& 1.2 & -0.21	& 0.18	& 2.35	& -12.25	& 1.0 & 0.00	& 0.01	& 5.25	& -12.40	& 1.9 & 0.12	& 0.20	\\
-14.44	& -8.08	& 6.2 & 0.01	& 0.00	& -15.54	& 6.54	& 1.2 & -0.20	& -0.03	& 13.40	& -4.28	& 0.9 & 0.00	& 0.01	& -21.68	& 7.00	& 1.9 & -0.11	& -0.02	\\
12.33	& -16.93	& 6.1 & 0.01	& 0.00	& 17.96	& 12.35	& 1.2 & 0.04	& 0.11	& -18.79	& 18.44 	& 1.0 & 0.00	& 0.01	& 9.33	& -15.35	& 1.9 & 0.10	& 0.18	\\
-17.84	& -0.29	& 6.4 & -0.01	& -0.01	& 8.87	& -14.92	& 1.2 & -0.12	& 0.08	& 10.53	& -17.81	& 1.0 & -0.01	& 0.00	& -3.35	& 2.22	& 1.9 & 0.38	& 0.14	\\
-11.19	& 7.51	& 6.2 & 0.01	& -0.00	& -13.77	& 0.49	& 1.4 & -0.28	& 0.02	& -0.92	& 16.80	& 1.0 & -0.01	& 0.00	& 0.15	& 7.75	& 2.1 & 0.08	& -0.15	\\
0.30		& 7.87	& 6.4 & 0.01	& -0.00	& 1.36	& 8.26	& 1.3 & 0.26	& 0.10	& 4.62	& -13.74	& 1.0 & -0.01	& 0.00	& -14.42 	& 6.90	& 1.9 & 0.36	& -0.09	\\
-11.93	& 2.29	& 6.1 & 0.00	& 0.01	& 5.41	& -17.85	& 1.2 & 0.26	& 0.07	& -4.68	& -18.59	& 1.0 & 0.01	& 0.00	& 7.94	& 15.33	& 1.8 & -0.14	& 0.18	\\
3.61		& 7.02	& 6.1 & -0.01	& 0.00	& 4.95	& 3.44	& 1.2 & 0.03	& 0.08	& 14.70	& -19.41	& 1.0 & -0.01	& 0.01	& 4.83	& -11.43	& 1.9 & -0.28	& 0.16	\\
9.76		& -7.40	& 6.1 & -0.01	& 0.01	& -14.24	& -4.71	& 1.2 & 0.08	& 0.02	& -14.63	& -11.18	& 1.0 & 0.01 	& 0.00	& 9.12	& -16.45	& 1.9 & 0.06	& 0.12	\\
\enddata
\label{table:1} 
\end{deluxetable}

\clearpage

\begin{deluxetable}{rrrrrrrrrrrrrrrrrrrr}
\tabletypesize{\tiny}
\addtolength{\tabcolsep}{-5pt}
\rotate
\tablecaption{A small sample of predicted guide corrections calculated off-sky for the $4$ instruments tested. X and Y are the required guide corrections in pixels between the reference image and a random sample of $10$ images from the observing block. X$_{\mathrm{r}}$ and Y$_{\mathrm{r}}$ are the residuals in pixels measured by DONUTS after the corrections have been applied manually in IRAF and t$_{\mathrm{calc}}$ is the time in ms of the calculation.}
\tablewidth{0pt}
\tablehead{
\multicolumn{5}{c}{INT - WFC} & \multicolumn{5}{c}{NGTS-P} & \multicolumn{5}{c}{WHT - AG2} & \multicolumn{5}{c}{NGTS-1}\\
\colhead{X} & \colhead{Y} & \colhead{t$_{\mathrm{calc}}$} & \colhead{X$_{\mathrm{r}}$} & \colhead{Y$_{\mathrm{r}}$} & \colhead{X} & \colhead{Y} & \colhead{t$_{\mathrm{calc}}$} & \colhead{X$_{\mathrm{r}}$} & \colhead{Y$_{\mathrm{r}}$} & \colhead{X} & \colhead{Y} & \colhead{t$_{\mathrm{calc}}$} & \colhead{X$_{\mathrm{r}}$} & \colhead{Y$_{\mathrm{r}}$} & \colhead{X} & \colhead{Y} & \colhead{t$_{\mathrm{calc}}$} & \colhead{X$_{\mathrm{r}}$} & \colhead{Y$_{\mathrm{r}}$} \\
\colhead{(pixels)} & \colhead{(pixels)} & \colhead{(ms)} & \colhead{(pixels)} & \colhead{(pixels)} & \colhead{(pixels)} & \colhead{(pixels)} & \colhead{(ms)} & \colhead{(pixels)} & \colhead{(pixels)} & \colhead{(pixels)} & \colhead{(pixels)} & \colhead{(ms)} & \colhead{(pixels)} & \colhead{(pixels)} & \colhead{(pixels)} & \colhead{(pixels)} & \colhead{(ms)} & \colhead{(pixels)} & \colhead{(pixels)} \\ 
}
\startdata
-0.01 & 0.73  &	13.0 & -0.01 &  0.01 & -0.00 & -0.03 & 4.3 & 0.01  & -0.02	&  -0.10 & -0.17  &  5.7 &  0.00 & -0.01 & 0.17  & -0.10  & 3.2 & 0.07 & -0.10   \\
-0.03	 & -5.34 &	12.2 & -0.02 &  0.01 & -0.02 & -0.04 & 3.6 & 0.01  & -0.03	&  -0.33 &  0.25   & 2.4 & -0.00 &  0.00 & 0.18  & -0.12  & 2.4 & 0.03 & -0.10  \\
-0.04	 & -7.72 &	12.0 & -0.03 &  0.01 & -0.03 & 0.00  & 3.2 & 0.02  &  0.00 &  -0.37 &  0.41   & 2.5 & -0.00 & -0.00 &  0.25  & -0.08 & 2.3 & 0.06 & -0.10  \\
-0.06	 & -7.38 &	11.9 & -0.04 &  0.00 & -0.01 & 0.03  & 4.1 & -0.00 &  0.01	&  -0.61 &  0.77   & 3.0 &  0.00 & -0.01 & 0.22  & -0.02  & 2.4 & 0.08 & -0.07   \\
-0.03	 & -6.29 &	11.9 & -0.02 & -0.01 & -0.06 & 0.03  & 3.2 & -0.01 &  0.02	& 0.46   &   0.14   & 2.4 &  0.01 &  0.01 & 0.21  & -0.10  & 2.3 & 0.06 & -0.09   \\
 0.02 & -4.43 &	12.0 &  0.02 & -0.01 & -0.06 & 0.07  & 3.3 & -0.04 &  0.04	&  0.42   &  -0.54 & 1.9 &  0.00 & -0.00 & 0.16  & -0.11  & 2.3 & 0.05 & -0.09   \\
 0.04 & -2.21 &	12.2 &  0.02 & -0.01 & -0.06 & 0.10  & 3.5 & -0.05 &  0.06	&  -0.13  & -0.06  & 2.4 &  0.00 &  0.00 & 0.15  & -0.08  & 2.4 & 0.05 & -0.06   \\
 0.04 & 0.30  &	12.8 &  0.02 &  0.02 & -0.09 & 0.14  & 3.2 & -0.04 &  0.07	&  0.04   &  0.64   & 2.1 & -0.00 & -0.01& 0.19  & -0.08  & 2.4 & 0.05 & -0.09   \\
 0.03 & 3.05  &	12.7 &  0.02 &  0.01 & -0.07 & 0.18  & 3.2 & -0.06 &  0.07	&  -0.03 &  0.08   & 2.5 & -0.00 &  0.00 & 0.14  & -0.01  & 2.3 & 0.08 & -0.04   \\
 0.02 & 1.46  &	14.7 &  0.01 &  0.05 & -0.08 & 0.19  & 3.5 & -0.09 &  0.07	&  -0.03 &  0.43   & 2.3 &  0.00 & -0.00 & 0.17  & -0.09  & 2.4 & 0.07 & -0.08   \\
\enddata
\label{table:2} 
\end{deluxetable}

\clearpage

\begin{deluxetable}{rl}
\tabletypesize{\scriptsize}
\addtolength{\tabcolsep}{-2pt}
\tablecaption{Assumed parameters used for calculating the limiting V magnitude in the worst case scenario of a single, low signal-to-noise guide star.}
\tablewidth{0pt}
\tablehead{
\colhead{Parameter} & \colhead{Assumed Value}\\
}
\startdata
Limiting SNR						& $15$ \\
Telescope aperture					& $1.0$ m \\
Camera platescale					& $0.5$ arcsec pixel$^{-1}$ \\
Read noise 						& $10$ e$^{-}$ \\
Dark current 						& $0.2$ e$^{-}$ s$^{-1}$ pixel$^{-1}$ \\
FWHM star						& $2.5$ pixels \\
Photometry aperture radius			& $2.5$ pixels \\
Total efficiency\tablenotemark{a} 		& $50$ \% \\
Vega flux							& $9.6\times10^{10}$ photons s$^{-1}$ m$^{-2}$ $\mu$m$^{-1}$ \\
$\lambda_{cen}$ V					& $0.555$ $\mu$m \\
$\lambda_{width}$ V				& $0.089$ $\mu$m \\
Sky level at full moon\tablenotemark{b}	& $18$ mag arcsec$^{-2}$ \\
Sky level at new moon\tablenotemark{b}	& $22$ mag arcsec$^{-2}$ \\
\enddata
\tablenotetext{a}{Includes atmospheric, telescope and filter throughputs plus CCD quantum efficiency in the V band.}
\tablenotetext{b}{We assume the sky brightness levels for La Palma listed at http://www.ing.iac.es/PR/newsletter/news6/tel1.html}
\label{table:3} 
\end{deluxetable}

\clearpage

\begin{deluxetable}{cccc}
\tabletypesize{\scriptsize}
\addtolength{\tabcolsep}{-2pt}
\tablecaption{Simulated seeing changes after applying a Gaussian blurring filter. Row $1$ represents the original data, X$_{\mathrm{r}}$ and Y$_{\mathrm{r}}$ are the average residuals for each $200$ image series at each simulated seeing level. $\delta_{\mathrm{FWHM}}$ is the percentage increase in FWHM at each stage with respect to the original data (row 1).}
\tablewidth{0pt}
\tablehead{
\colhead{FWHM} & \colhead{X$_{\mathrm{r}}$} & \colhead{Y$_{\mathrm{r}}$} & \colhead{$\delta_{\mathrm{FWHM}}$}\\
\colhead{(pixels)} & \colhead{(pixels)} & \colhead{(pixels)} & \% \\
}
\startdata
1.32	& 0.18	& 0.11	& ---   \\
1.58	& 0.18	& 0.10	& 20  \\
1.78	& 0.18	& 0.10 	& 35 \\
1.98	& 0.19	& 0.10	& 50   \\
2.18	& 0.20	& 0.09	& 65   \\
2.37	& 0.20	& 0.09	& 80   \\
2.58	& 0.21	& 0.09	& 95   \\
2.78	& 0.22	& 0.09	& 110   \\
2.98	& 0.23	& 0.09	& 125  \\
\enddata
\label{table:4} 
\end{deluxetable}

\clearpage


\clearpage

\clearpage

\clearpage

\clearpage

\clearpage

\clearpage

\clearpage


\end{document}